\documentclass[draft]{agujournal2019}
\usepackage{url} 
\usepackage{lineno}
\usepackage[inline]{trackchanges} 
\usepackage{ragged2e}
\usepackage{soul}
\usepackage{amsthm}
\usepackage{amssymb}
\usepackage{amsmath}
\usepackage{bm}
\usepackage{tabularray}
\usepackage{amsmath} 
\usepackage{xcolor,colortbl}
\usepackage{colortbl}
\usepackage{float}
\usepackage{booktabs}
\usepackage{soul}

\floatstyle{plaintop}
\restylefloat{table}

\newcommand{\corr}[1]{#1}

\definecolor{Gray}{gray}{0.85}
\definecolor{LightCyan}{rgb}{0.88,1,1}

\newcolumntype{a}{>{\columncolor{Gray}}c}
\newcolumntype{b}{>{\columncolor{white}}c}

\draftfalse

\journalname{JGR: Space Physics}

\justifying
\begin{document}


\newcommand{\sd}[1]{{\color{cyan}{#1}}}
\newcommand{\sdc}[1]{{\color{blue}{Sc: #1}}}
\newcommand{\p}[1]{{\color{magenta}{#1}}}
\newcommand{\pc}[1]{{\color{green}{(P comment: #1)}}}
\newcommand{\pr}[1]{{\color{yellow}{(P remove: #1)}}}
\newcommand{\vl}[1]{{\color{darkgreen}{#1}}}
\newcommand{\vlc}[1]{{\color{orange}{(Vc: #1)}}}
\newcommand{\oa}[1]{#1}
\newcommand{\sv}[1]{{\color{violet}{#1}}} 
\newcommand{\svp}[1]{{\color{blue}{#1}}} 
\newcommand{\svtodo}[1]{{\color{red}{#1}}} 
\newcommand{\correction}[1]{{#1}}

\newcommand{\BE}{\begin{equation}}
\newcommand{\EE}{\end{equation}}
\newcommand{\figsss}[1]{Figure~\ref{fig_#1}}
  \newcommand{\eqp}[1]{(Eq.~\ref{eq_#1})}
 \newcommand{\eqs}[2]{Eqs.~(\ref{eq_#1}) and (\ref{eq_#2})}
 \newcommand{\eqss}[2]{Eqs.~(\ref{eq_#1}) - (\ref{eq_#2})}
 \newcommand{\eqsss}[3]{Eqs.~(\ref{eq_#1}), (\ref{eq_#2}) and (\ref{eq_#3})}
\newcommand{\eg}{{\it e.g.}}
\newcommand{\etal}{{\it et al.}}
\newcommand{\ie}{{\it i.e.}}
\newcommand{\insitu}{{\it in situ }}
\newcommand{\cf}{{\it cf.}}
\newcommand{\tab}[1]{Table~\ref{tab_#1}}

\newcommand{\f}[2]{{\ensuremath{\mathchoice%
        {\dfrac{#1}{#2}}
        {\dfrac{#1}{#2}}
        {\frac{#1}{#2}}
        {\frac{#1}{#2}}
        }}}
\newcommand{\Int}[2]{\ensuremath{\mathchoice%
        {\displaystyle\int_{#1}^{#2}}
        {\displaystyle\int_{#1}^{#2}}
        {\int_{#1}^{#2}}
        {\int_{#1}^{#2}}
        }}

\newcommand{\curl}{ {\bf \nabla} \times}
\newcommand{\D}{\partial}
\renewcommand{\div}[1]{ {\bf \nabla }. #1 }
\newcommand{\grad}{ {\bf \nabla } }
\newcommand{\pder}[2]{\f{\partial #1}{\partial #2}}
\newcommand{\lapO}{ {\nabla^2_\bot } }
\newcommand{\pxy}[2]{\{#1,#2 \}_{x,y}}
\newcommand{\pxeta}[2]{\{#1,#2 \}_{x,\eta}}
\newcommand{\rmd}{{\rm d }}
\newcommand{\tder}[2]{\f{D #1}{D #2}}
\newcommand{\X}{\times}

\newcommand{\uvec}[1]{\hat{\bf #1}}
\newcommand{\tA}{\tilde{A}}
\newcommand{\tPhi}{\tilde{\Phi}}

\newcommand{\vA}{\mathbf{A}}
\newcommand{\vB}{\mathbf{B}}
\newcommand{\vb}{\hat{\mathbf{b}}}
\newcommand{\vj}{\mathbf{j}}
\newcommand{\vv}{\mathbf{v}}
\newcommand{\vBo}{\mathbf{B}_0}
\newcommand{\vV}{\mathbf{V}}
\newcommand{\vU}{\mathbf{U}}

\newcommand{\eone}{\mathbf{e}_1}
\newcommand{\etwo}{\mathbf{e}_2}
\newcommand{\ethr}{\mathbf{e}_3}
\newcommand{\ve}{\mathbf{e}}
\newcommand{\perpLAPLAS}{\nabla_{\perp}^{2}}
\newcommand{\perpLAPLASt}{\tilde{\nabla}_{\perp}^{2}}
\newcommand{\vardbtilde}[1]{\tilde{\raisebox{0pt}[0.85\height]{$\tilde{#1}$}}}

\newcommand{\au}{\rm au}
\newcommand{\degree}{\ensuremath{^\circ}}
\newcommand{\kms}{\rm km~s$^{-1}$}


\newcommand{\eps}{    {Earth, Planets, and Space}}
\newcommand{\adv}{    {Adv. Spa. Res.}}
\newcommand{\annG}{   {Annales Geophysicae}}
\newcommand{\ag}{   {Annales Geophysicae}}
\newcommand{\aap}{    {\it Astron. Astrophys.}}
\newcommand{\aapr}{   {\it Astron. Astrophys. Rev.}}
\newcommand{\apj}{    {\it Astrophys. J.}}
\newcommand{\jastp}{  {J. Atmos. Sol. Terr. Phys.}}
\newcommand{\jfm}{J. Fluid. Mech. }
\newcommand{\jgr}{    {\it J. Geophys. Res.}}
\newcommand{\lrsp}{    {\it Living Rev. Sol. Phys.}}
\newcommand{\mnras}{  {\it Mon. Not. Roy. Astron. Soc.}}
\newcommand{\natc}{    {\it Nature Com.}}
\newcommand{\solphys}{{\it Solar Phys.}}


\newcommand{\vdag}{(v)^\dagger}
\newcommand\aastex{AAS\TeX}
\newcommand\latex{La\TeX}

\title{Kinetic-based macro-modeling of the solar wind at large heliocentric distances: Kappa electrons at the exobase}

\authors{A. Vinogradov\affil{1}\thanks{Correspondence to: alexander.vinogradov@kuleuven.be},
M. Lazar\affil{1,2}, 
I. Zouganelis\affil{3}, 
V. Pierrard\affil{4,5}, 
S. Poedts\affil{1,6}}

\affiliation{1}{Centre for Mathematical Plasma-Astrophysics, KU Leuven, Celestijnenlaan 200B, 3001 Leuven, Belgium}
\affiliation{2}{Theoretical Physics IV, Faculty for Physics and Astronomy, Ruhr-University Bochum, Bochum, Germany}
\affiliation{3}{European Space Agency (ESA), European Space Astronomy Centre (ESAC), Camino Bajo del Castillo s/n, 28692 Villanueva de la Cañada, Madrid, Spain}
\affiliation{4}{Royal Belgian Institute for Space Aeronomy (BIRA-IASB), Space Physics, Solar Terrestrial Center of Excellence (STCE), Brussels B-1180, Belgium}
\affiliation{5}{Earth and Life Institute-Climate Sciences (ELI-C), Université Catholique de Louvain, Louvain-la-Neuve, Belgium}

\affiliation{6}{Institute of Physics, University of Maria Curie-Skłodowska, Pl. M. Curie-Skłodowska 5, 20-031 Lublin, Poland}


\begin{keypoints}
\item Suprathermal electrons with Kappa distributions (KDs) in the solar corona play a key role in the formation/acceleration of the solar wind.
\item A semi-analytic kinetic model is built for the large-distance solar wind by assuming electrons with regularized KDs (RKDs) at the exobase.
\item RKDs resolve the existing inconsistencies and reduce possible overestimations from the increased presence of suprathermal electrons.
\end{keypoints}

\abstract{Recent evidence from Parker Solar \corr{Probe} on the suprathermal electrons with Kappa-type velocity distributions in the outer corona has revived interest in the kinetic-based macro-modelling of the solar (SW), aiming to explain its properties. 
Invoked in kinetic modelling of nonequilibrium plasmas, standard Kappa distributions (SKDs) have been adjusted to the regularized Kappa distributions (RKDs) to fix the inconsistencies of SKD and develop consistent fluid modelling of space plasmas.
We propose a new analysis of these properties at large heliocentric distances based on the existence of RKD electrons at the exobase. 
This new semi-analytic formalism is inspired by the methodology proposed initially by \citeA{MeyerVernet-Issautier-1998}. 
Compared to SKDs, the results for RKDs have extended applicability, since all moments can be defined and calculated consistently for all values of the $\kappa$ parameter, even lower than the critical ones (e.g., $\kappa_c=3/2$ imposed to the second-order moment) of SKDs. 
However, the excess energy of the more energetic suprathermal electrons associated with low values of $\kappa \lesssim 3/2$, is regulated by the RKD-specific cutoff parameter $\alpha < 1$.
The estimates for, e.g., the temperature and bulk velocity of the SW, remain at realistic values even for small $3/2 < \kappa \lesssim 2$, which would otherwise exceed specific observations.
One can thus model a higher abundance of suprathermal electrons at the exobase (e.g., $\kappa \leqslant 3/2$), which is plausible for the sources of energetic events (flares and coronal mass ejections), and also in the astrospheres of stars with coronas hotter than the Sun's.
}

\section*{Plain Language Summary}
The solar wind is a flow of charged particles, electrons and ions/protons, emerging from the outer solar corona and reaching speeds between 300 and 800 km/s in interplanetary space.
Kinetic-exospheric models show that the solar wind can be driven by light but energetic, suprathermal electrons that escape solar gravity and create an electric field that pulls protons away from the Sun. 
We provide an improved semi-analytic model of the long-distance solar wind \corr{($r > 0.3\,\mathrm{AU}$)} that uses a regularized Kappa distribution to realistically capture the effects of suprathermal electrons at the exobase of a few solar radii. 
This distribution resolves the inconsistencies and limitations of standard Kappa distributions, offering a consistent extension of such macro-models, even in the presence of enhanced suprathermal electrons at the exobase, characteristic of more energetic coronal outflows.
%

\section{\bf Introduction} 

The ability of kinetic exospheric models to produce realistic estimates of the solar wind (SW) properties depends on how good the assumptions made for the initial or boundary conditions are, especially those at the exobase (a few solar radii), the altitude above which there are no collisions \cite{Lemaire-Scherer-1971, Maksimovic_etal_1997, Pierrard-etal-2001, Lamy-etal-2003, Zouganelis-etal-2004, Vocks-etal-2005, Pierrard-etal-2011, Peter-de-Bonhome-etal-2025}. 
While these estimates have been able to be evaluated for decades by comparing them with the SW properties measured since the first space missions at different heliocentric distances (such as Helios, Ulysses and those orbiting Earth), the scenarios proposed for the exobase are only now being confirmed (or not) thanks to coronal observations by the Parker Solar Probe [PSP] \cite{Abraham-etal-2022, Zheng-etal-2024}.

A key ingredient in the formation and acceleration of the SW is energetic suprathermal electrons, which are not in thermal equilibrium and can be described by Kappa ($\kappa$-)power-law velocity distributions \cite{Pierrard-Lazar-2010, Lazar-Fichtner-2021}.
Kappa electrons are ubiquitous in the SW \cite{Maksimovic-etal-1997b, Stverak-etal-2008, Wilson-etal-2019}, and PSP data also support their presence in the outer corona \cite{Abraham-etal-2022, Zheng-etal-2024}.
Suprathermal electrons enhance the number of electrons escaping at the exobase and implicitly determine an increase in the ambipolar electrostatic (ES) potential that drags protons (overcoming gravitational attraction) and accelerates them to supersonic speeds \cite{Lamy-etal-2003, Zouganelis-etal-2004}.
Moreover, lower exobases lead to increased bulk velocities of the SW \cite{Zouganelis-etal-2004}, in agreement with the (hypothetical) subcoronal origin of fast outflows, commonly linked to coronal holes. 
\corr{ 
Thus, exospheric models consider the exobase at small distances, below 10~$R_S$, assuming that beyond the exobase particle-particle collisions are no longer effective in establishing thermodynamic equilibrium \cite{MeyerVernet-Issautier-1998, Lamy-etal-2003, Zouganelis-etal-2004}.
Recent in situ observations of PSP down to 10~$R_S$ support these scenarios by revealing electrons with Kappa (non-Maxwellian) VDs, with $\kappa$ reaching low values around 4, and even lower, as suggested by the decreasing radial trend \cite{Abraham-etal-2022, Zheng-etal-2024}.
There are also indirect indications from modelling supporting the presence of such energetic electron populations in the solar corona, whose inclusion enhances both heavy ion charge states and solar wind acceleration, improving agreement with observations \cite{Lomazzi-etal-2025, Shen-etal-2025}.
An exobase at low radial distance like $6 R_S$ \cite{Lemaire-Scherer-1971, MeyerVernet-Issautier-1998} can be justified not only by the collisional rate, but also the configuration of the magnetic field opening to the interplanetary space at this distance, which can explain why the slow wind is characterized by a higher exobase than the fast wind \cite{Pierrard-etal-2023}.
}

The values of the parameter $\kappa$ measured for the electron \corr{velocity distributions (VDs hereafter)} in high-speed SW are lower (the suprathermal tails are stronger) than in slow winds \cite{Maksimovic-etal-1997b, Lazar-etal-2020}, also suggesting a relationship between the suprathermal electrons and the bulk speed.
The self-consistent evaluation of the ES potential at each position in space is not straightforward, being generally achieved by iterative calculations based on the neutrality of the electric charge and current, sufficiently well preserved in the SW \cite{Feldman-etal-1975}.
An analytical or semi-analytical solution of this ES potential is thus of great interest, as it can significantly facilitate calculation even in numerical codes of exospheric models \cite{MeyerVernet-Issautier-1998, Osuna-etal-2021}.

Despite the progress in describing the VDs observed in non-equilibrium plasmas in space, standard Kappa distributions (SKDs) still have inconsistencies that have been resolved more recently by the so-called regularized Kappa distributions (RKDs) \cite{Scherer-etal-2017, Lazar-Fichtner-2021}.
The additional exponential factor introduces the new cutoff parameter $0< \alpha < 1$, which primarily reduces the unrealistic contribution of superluminal (nonphysical) particles with velocities $v>c$ (where $c$ is the speed of light in vacuum).
Furthermore, important for large-scale (macro) models, such as those pursued by exospheric kinetic modelling, is that all moments of RKDs are well defined for all values of the parameter $\kappa > 0$.
For SKDs, moments of order $l \geqslant 2$ and the corresponding transport coefficients take finite values only for large enough $\kappa > (l + 1)/2$, which is not the case for RKDs \cite{Lazar-etal-2020b, Husidic-etal-2022}.
The capabilities to describe suprathermal populations and their implications are consistently expanded. 
Numerical solutions of a preliminary model constructed for RKD electrons at the exobase have been reported, limiting, however, to large values of $\kappa > (l + 1)/2$ for easier comparison with previous results obtained for SKD electrons \cite{Pierrard-etal-2023}.
Therefore, RKDs also offer an improved alternative for describing VDs observed in situ, without having to disregard those with strong suprathermal tails, associated with a reduced parameter $\kappa$ \cite{Scherer-etal-2021}. 

This paper proposes a new kinetic modeling of SW properties at large heliocentric distances from the exobase, following the semi-analytic formalism of \citeA{MeyerVernet-Issautier-1998}.
These properties depend on the nature of the VDs of the electron and proton populations at the exobase. 
This already follows from the general expressions for the SW parameters, derived previously (for arbitrary VDs) at large distances, and briefly reviewed in Section~2.
Even at large heliocentric distances, up to 1~AU and beyond, where the variation with distance remains roughly constant, the absolute values of the SW parameters may significantly depend on the nature of the electron VDs at the exobase \cite{MeyerVernet-Issautier-1998}.
To be sufficiently explicit, Section~3 presents a retrospective of the previous approach based on the presence of SKD electrons at the exobase, as well as a new extended analysis that provides analytic solutions that accurately reproduce the exact ones.
The new semi-analytic model, assuming RKD electrons at the exobase, is provided in Section~4.
The complexity of the solutions obtained in this case for the ES potential and the SW parameters, expressed in terms of hypergeometric Tricomi functions, precludes rigorous approximations and simplified, fully analytical forms.
However, kinetic models based on RKD electrons offer opportunities to analyze new conditions at the exobase, with suprathermal electrons of increased densities and energies that SKDs cannot describe. 
These could thus explain the coronal outflows with bulk velocities much higher than those of the SW associated with energetic solar events, but can also contribute to modeling the atmospheres and winds of stars with coronas hotter than the Sun.
Such perspectives are discussed, along with the results of this work, in Section~5.

\section{\bf General approach}\label{sec2}
Following the general formalism in \citeA{MeyerVernet-Issautier-1998}, we present here only the main relationships and parameters used to determine the ES potential at the exobase, the terminal SW speed, and the heliocentric radial profiles at large distances. 
At this stage, these are independent of any specific form of electron or proton VDs, but will be specialized to SKD electrons in section~\ref{sec3} and to RKD electrons in section~\ref{sec4}.
General expressions of the moments for the (arbitrary) distribution functions of protons and electrons are presented in \ref{secA} and \ref{secB}, respectively.
For additional details that may aid in understanding these results, we recommend consulting the basic formalism in \citeA{MeyerVernet-Issautier-1998}.

\subsection{Boundary conditions at the exobase}\label{sec:2_1}
To evaluate the ES potential at the exobase level, we enforce the condition of zero net flux of charge, which is given by the equality of electron (subscript $e$) and proton (subscript $p$) fluxes \cite{MeyerVernet-Issautier-1998}
\begin{align}\label{e1}
    F_e \equiv \int_{V_0}^{\infty}dv v^3 f_{e0}(v)= F_p \equiv \int_{0}^{\infty}dv v^3 f_{p0}(v),
\end{align}
where $f_{e0}$ and $f_{p0}$ are electron and proton distribution functions at the exobase level $r=r_0$. 
The parameter $V_0$, introduced in \citeA{MeyerVernet-Issautier-1998}, corresponds to the characteristic electron escape speed; only electrons with $v>V_0$ contribute to the escaping flux. 
It is related to the ES potential $\Phi_E(r_0)$ through equation (9) in \citeA{MeyerVernet-Issautier-1998},
\begin{align}\label{e2}
V_0 = \sqrt{\frac{2 e \Phi_E(r_0)}{m_e}} .
\end{align}
The distribution functions $f_{p0}(v)$ and $f_{e0}(v)$ are normalized according to equations (18) and (19) of \citeA{MeyerVernet-Issautier-1998}
\begin{align}
    n_e(r_0) &= 4\pi \int_{0}^{\infty} dv \, v^2 f_{e0}(v) \;-\; 2\pi \int_{V_0}^{\infty} dv \, v^2 f_{e0}(v),\label{e3} \\
    n_p(r_0) &= 2\pi \int_{0}^{\infty} dv \, v^2 f_{p0}(v).\label{e4}
\end{align}
Once normalizations are determined, we substitute $f_{p0}(v)$ and $f_{e0}(v)$ into \eqref{e1}, use the quasi-neutrality condition $n_e(r_0) = n_p(r_0)$, and solve equation~\eqref{e1} for $\Phi_E(r_0)$. 
This boundary value of the potential allows us to estimate the terminal SW speed $V_{\mathrm{SW}}$ from the approximate proton energy balance $m_p V_{\mathrm{SW}}^2/2 \approx e\Phi_E(r_0) - m_p \Phi_G(r_0)$ \cite{MeyerVernet-Issautier-1998}, as
\begin{align}\label{e5}
V_{\mathrm{SW}} \approx \{2[e\Phi_E(r_0) - m_p \Phi_G(r_0)]/m_p\}^{1/2},
\end{align}
with the gravitational potential at the exobase given by
\begin{align}\label{e6}
    \Phi_G(r_0)=\frac{GM_{S}}{r_0}.
\end{align}
%

\subsection{Radial profiles at large distances} \label{sec:2_2}
We adopt the asymptotic expressions for radial profiles derived by \citeA{MeyerVernet-Issautier-1998}. These expressions are valid in the limit of large distances, where $\eta(r) = \left( r_0 / r \right)^2 \ll 1$. 
The radial profile of the density is given by
\begin{align} \label{e7}
    n_p(r)=C_0 \eta(r)=C_0\left(\frac{r_0}{r}\right)^2,
\end{align}
where
\begin{align} \label{e8}
    C_0=\pi \int_{0}^{\infty} dv \frac{v^3}{\sqrt{(v^2+V_{p0}^2)}}f_{p0}(v),
\end{align}
and
\begin{align} \label{e9}
V_{p0} = \sqrt{\frac{2}{m_p}(e\Phi_E(r_0)-\Phi_G(r_0))}.
\end{align}
The radial profile of the ES potential is given by 
\begin{align}\label{e10}
    \Phi_E(r) = \frac{m_e}{2 e} \left( \frac{C_0}{D_0} \right)^{2/3} \eta^{2/3} =  \frac{m_e}{2 e} \left( \frac{C_0}{D_0} \right)^{2/3} \left(r_0/r\right)^{4/3},
\end{align}
where 
\begin{align}\label{e11}
    D_0=\frac{4\pi }{3} f_{e0}(V_0),
\end{align}
depends on the electron VD. 
The electron temperature can be expressed as a sum of two terms 
\begin{align} \label{e12}
    T_e (r)&=T_{4/3}(r) + T_C.
\end{align}
The first term corresponds to the effect of nonescaping electrons and depends explicitly on $r$
\begin{align} \label{e13}
T_{4/3} (r)&=\frac{m_e}{5k_B}\left(\frac{C_0}{D_0}\right)^{2/3}\eta^{2/3}\propto \left(\frac{r_0}{r}\right)^{4/3}.
\end{align}
The second term, $T_C$, arises from escaping electrons. It is independent of the radial distance $r$ and is given by 
\begin{align} \label{e14}
T_C &= \frac{m_e \, B_2}{3 k_B C_0},
\end{align}
where
\begin{align} \label{e15}
    B_2=\pi\int_{V_0}^{\infty}dv v^3 \sqrt{v^2-V_0^2} f_{e0} (v).
\end{align}
These relations are the general framework of our study. In the following sections, we apply this procedure to specific cases: SKD electrons with Maxwellian protons (Sec.~\ref{sec3}), and then RKD electrons (Sec.~\ref{sec4}). 

\section{\bf SKD electrons and Maxwellian protons at the exobase} \label{sec3}
The proton VD at the exobase is assumed to be Maxwellian:
\begin{align}
    \label{e16}
    f_{0p}=A_p\exp\left(\frac{v^2}{w_{p0}^2}\right),
\end{align}
and the normalization constant $A_p$ is derived from \eqref{e4}
\begin{align}
    \label{e17}
    A_p=\frac{2n_{p}(r_0)}{(\pi w_{p0}^2)^{3/2}}.
\end{align}
In the previous approach, VD models were introduced with normalization constants, such as generic densities, e.g., $n_{p0}$ for protons and $n_{e0}$ for electrons \cite{MeyerVernet-Issautier-1998}. 
\corr{ Here we show that (non-normalized) physical quantities at the exobase can also be used, such as for the densities $n_{p}(r_0)$ and $n_{e}(r_0)$, as well as thermal velocities $w_{p0}$ and $w_{e0}$.}
We first revisit the approach in \citeA{MeyerVernet-Issautier-1998}, where the electrons follow an SKD
\begin{align}\label{e18}
    f_{0e}=A_e\left(1+\frac{v^2}{\kappa w_{e0}^2}\right)^{-\kappa-1}.
\end{align}
The expression in \eqref{e3} is then used to find the normalization constant
\begin{align} \label{e19} 
A_{e}=\frac{n_{e}(r_0)}{(\pi\kappa w_{e0}^{2})^{3/2}}\frac{1}{\left[ \frac{1}{2}\frac{\Gamma(\kappa-1/2)}{\Gamma(\kappa+1)}+\frac{B_{x_0}(3/2, \kappa-1/2)}{\sqrt{\pi}}\right]},
\end{align}
where 
\begin{align} \label{e20}
  B_{x_0} (a,b)= \int_{0}^{x_0} t^{a-1}(1-t)^{b-1}dt,
\end{align}
is the incomplete beta function with 
\begin{align}\label{e21}
x_0=\frac{V_0^2}{\kappa w_{e0}^2+V_0^2} = {y \over y + 1} < 1,
\end{align}
and the key parameter
\begin{align} \label{e22}
y=\frac{V_0^2}{\kappa w_{e0}^2}.
\end{align}
The equivalence between our normalization for the Kappa distribution and that in \citeA{MeyerVernet-Issautier-1998} is proven in \ref{secC}. 
The ES potential at the exobase $\Phi_E(r_0)$ is then derived, according to its definition, from \eqref{e22}  
\begin{align}\label{e23}
\Phi_E(r_0) \equiv {m_e V_0^2 \over 2 e} = {m_e \kappa w_{e0}^2 \over 2 e} y.
\end{align}
%

\subsection{Exact (numerical) solutions} \label{sec3_1}

Using \eqref{e16}-\eqref{e19} and charge neutrality at $r_0$, $n_{e}(r_0) = n_{p}(r_0)$, the zero-net current condition in~\eqref{e1} gives us
\begin{align}\label{e24}
\frac{w_{p0}}{w_{e0}}= \frac{1+\kappa y }{\sqrt{\kappa}(\kappa-1)} \; \frac{(1+y)^{-\kappa}}{\frac{1}{2}\frac{\Gamma(\kappa-1/2)}{\Gamma(\kappa+1)}+\frac{B_{x_0}(3/2, \kappa-1/2)}{\sqrt{\pi}}}.
\end{align}
We numerically solve \eqref{e24} for $y$, using the Brent root-finding method \cite{Brent1973}. 
Once $y$ is computed, we then derive the ES potential at the exobase $\Phi_E(r_0)$ and the terminal SW speed $V_{\mathrm{SW}}$, using \eqref{e23} and \eqref{e5} respectively. 
We assume $r_0=6R_S$, as in \citeA{Lemaire-Scherer-1971} and \citeA{MeyerVernet-Issautier-1998}, where $R_S$ is the solar radius.
This is used to evaluate the gravitational potential defined in \eqref{e6}, which then enters into \eqref{e5}. 

Table~\ref{tab1} shows the exact numerical solutions $y$ for a range of $\kappa$ values, together with the parameters $\Phi_E(r_0)$ and $V_{SW}$ derived as described above. 
\corr{ Calculations are performed for standard coronal values at the exobase, namely, for a proton thermal speed $w_{p0} = 182$ km/s, corresponding to $T_{p0}= m_p w_{p0}^2 /2k_B=2 \times 10^6$ K,  and an electron thermal speed $w_{e0}=5.5 \times 10^3$ km/s, corresponding to $T_{e0}=m_ew_{e0}^2/2k_B=10^6$ K \cite{MeyerVernet-Issautier-1998}.}
These are typical values of proton and electron temperatures in the corona, in this case associated with the exobase at $r_0 = 6R_S$. 
According to Table~\ref{tab1}, the ES potential $\Phi_E(r_0)$ is lower for larger $\kappa$, while the SW is slower. This is a well-known effect \cite{Maksimovic_etal_1997} because a larger $\kappa$ means the electron distribution is closer to a Maxwellian, with a weaker high-energy tail. Fewer escaping electrons require a smaller confining potential, which in turn provides less acceleration for the SW.

\begin{table} 
\centering
\caption{Values obtained for the exact solutions of $y$, $\Phi_E(r_0)$, and $V_{SW}$ for $w_{p0} = 182$ km/s ($T_{p0}=2 \times 10^{6}$ K)  and $w_{e0}=5.5\times 10^3$ km/s ($T_{e0}= 10^{6}$ K), and different values $\kappa$.}
\label{tab1}
\begin{tabular}{c c c c}
\hline\hline
$\kappa$ & $y$ & $\Phi_E(r_0)$ [V] & $V_{SW}$ [km/s] \\
\hline
1.6& $2011.4$ & $2.77\times10^{5}$ & $7285$ \\
2  & 95.3 & $1.642\times10^{4}$ & $1756$ \\
3  & 9.5  & $2.468\times10^{3}$ & $640$ \\
5  & 2.4  & $1.036\times10^{3}$ & $367$ \\
10 & 0.8  & $6.515\times10^{2}$ & $247$ \\
20 & 0.3  & $5.375\times10^{2}$ & $198$ \\
\hline
\end{tabular}

\end{table}

\subsection{Analytical approximations}\label{sec3_2}

As argued in \citeA{MeyerVernet-Issautier-1998}, $V_0$ is generally much larger than the most probable speed of the electron distribution $f_{e0} (v)$, such that $V_0 >> w_{e0}$, while the contribution of the second term in \eqref{e3} is therefore negligible. 
In this limit, if $\kappa$ is small enough, we can assume $y= V_0^2/ (\kappa w_{e0}^2) >> 1$, and find $x_0 \to 1$, while our $B_{x_0}(3/2,\kappa-1/2)$ reduces to a constant 
\begin{align}
    \label{e25}
    \lim_{x_0\to1}B_{x_0}(3/2,\kappa-1/2)=\frac{\sqrt{\pi}}{2}\frac{\Gamma(\kappa-1/2)}{\Gamma(\kappa+1)}.
\end{align}
Using \eqref{e25},
\eqref{e24} reduces to 
\begin{align}\label{e26}
\frac{w_{p0}}{w_{e0}}=\frac{(1+\kappa y)(1+y)^{-\kappa} }{\sqrt{\kappa}(\kappa-1)}\frac{\Gamma(\kappa+1)}{\Gamma(\kappa-1/2)},
\end{align}
which can restrain to
\begin{align} \label{e27}
\kappa y + 1 = a (y + 1)^\kappa ,
\end{align}
with
\begin{align} \label{e28}
    a=\frac{w_{p0}}{w_{e0}}\frac{\Gamma(\kappa-1/2)}{\sqrt{\kappa}\Gamma(\kappa-1)}.
\end{align}
In the same limit of large $y_0 >> 1$, also invoked in \citeA{MeyerVernet-Issautier-1998}, we obtain an analytical zero-order approximation,
\begin{align}\label{e29}
    y_0 \simeq (\kappa/a)^{1/(\kappa-1)}, 
\end{align}
and the corresponding expression of the ES potential follows from \eqref{e23}
\begin{align}\label{e30}
\Phi_E(r_0) = {m_e w_{e0}^2 \over 2 e} \left({\kappa^\kappa} \over a \right)^{1\over \kappa-1}.
\end{align}
This result was first obtained in \citeA{MeyerVernet-Issautier-1998} in their equation (73), taking into account their definition (58).
However, the zero-order approximation is not accurate for large $\kappa$, as shown here in Figure~\ref{fig1}. 
Therefore, in the following, we derive a first-order approximation, which extends the validity of the analytic solution for larger $\kappa$. 
We introduce the correction term $\delta y$, such that
\begin{equation}\label{e31} 
    y \simeq y_0 + \delta y.
\end{equation}
Substituting \eqref{e31} into \eqref{e27} gives
\begin{align}\label{e32} 
1 + \kappa (y_0 + \delta y) = a \left(1 + y_0 + \delta y \right)^\kappa,
\end{align}
which can be rewritten as
\begin{align}\label{e33} 
    1 + \kappa (y_0 + \delta y) = a y_0^{\kappa} \left(1 + \frac{1 + \delta y}{y_0} \right)^\kappa.
\end{align}
%
For large $y_0$, we assume that the correction $\delta y$ is small, i.e. $(1+\delta y)/y_0 \ll 1$. Expanding to first order gives
\begin{align}\label{e34} 
\left(1 + \frac{1 + \delta y}{y_0}\right)^\kappa \approx 1 + \frac{\kappa (1 + \delta y)}{y_0} .
\end{align}
Substituting this expansion into \corr{\eqref{e33}} yields
\begin{align} \label{e35} 
1 + \kappa (y_0 + \delta y) &= a y_0^{\kappa} \left[1 + \frac{\kappa (1 + \delta y)}{y_0}\right].
\end{align}
From \eqref{e22}, we have $y_0^{\kappa-1}=\kappa/a$, and multiplying both sides by $ay_0$ gives $a y_0^{\kappa} = \kappa y_0$, which is then substituted in \eqref{e35} to find 
\begin{align}\label{e36} 
1 + \kappa (y_0 + \delta y) = \kappa y_0 \left(1 + \frac{\kappa (1 + \delta y)}{y_0}\right),
\end{align}
which simplifies, 
\begin{align}\label{e37} 
1 + \kappa \delta y = \kappa^2 (1 + \delta y), 
\end{align}
to finally obtain
\begin{align}\label{e38} 
    \delta y = -\frac{\kappa + 1}{\kappa} .
\end{align}
Substituting \eqref{e29} and \eqref{e38} into \eqref{e31} leads to the analytical expression for the first-order approximation of $y$,
\begin{align}\label{e39} 
y = \left(\frac{\kappa}{a}\right)^{\frac{1}{\kappa-1}}-\frac{\kappa + 1}{\kappa} .
\end{align}

\subsection{Exact solutions vs. analytical approximations}\label{sec3_4}

\begin{figure}[t!]
        \centering
\includegraphics[width=1\linewidth]{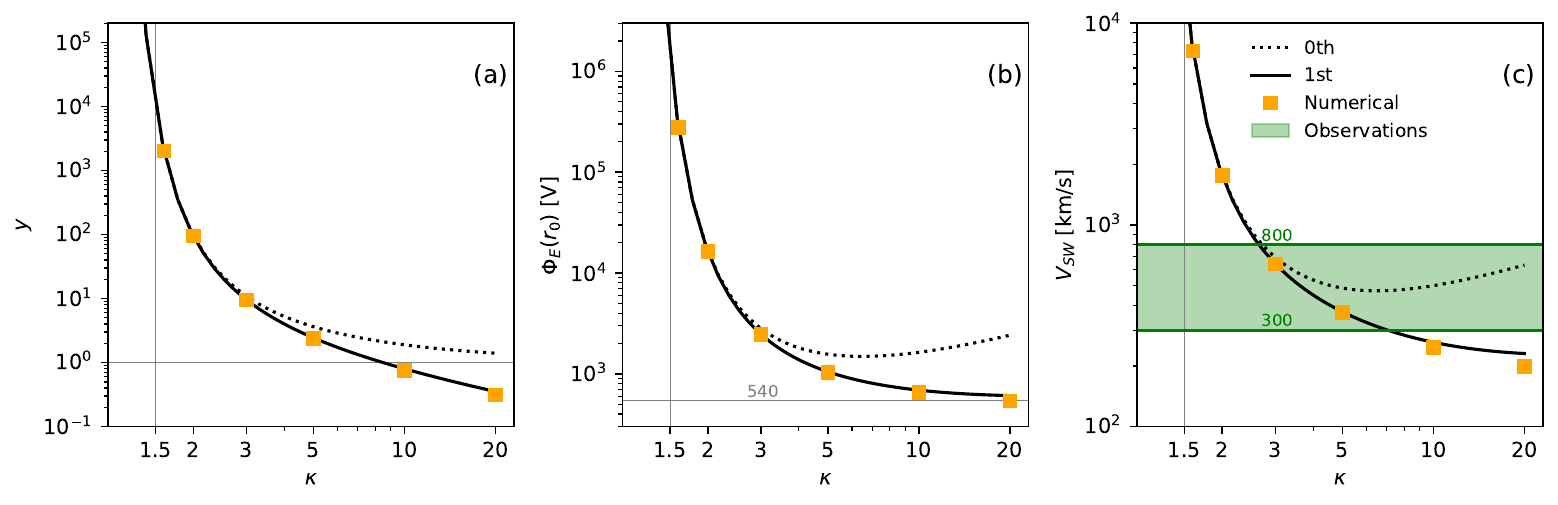}
\caption{Dependence of the solutions at the exobase on $\kappa$ for SKD electrons: (a) parameter $y$; (b) ES potential (dashed line marks the Maxwellian limit, $\kappa \rightarrow \infty$, when $\Phi_{E}(r_{0}) \simeq 540$ V); (c)~asymptotic (terminal) SW speed $V_{\mathrm{SW}}$. Plotted are the zero-order approximations (dotted lines), the first-order approximations (solid lines), and the exact numerical results (orange squares). The greenish range in panel (c) delimits the observed limits of the SW speed (300–800 km/s).} \label{fig1}
\end{figure}

In Figure~\ref{fig1}, we compare analytical approximations of zeroth order (dotted lines) and first order (solid lines) with the exact numerical solutions from Table~\ref{tab1} (orange squares), as obtained for different values of $\kappa$. 
The model parameters are set by assuming the exobase altitude at $r_0 = 6R_S$, a proton thermal speed of $w_{p0} = 182$ km/s (corresponding to $T_p = 10^6$~K), and an electron thermal speed of $w_{e0} = 5.5 \times 10^3$ km/s (corresponding to $T_e = 2 \times 10^6$~K).
The results are limited to $\kappa>3/2$, the validity range for the SKD, to keep the electron temperature physically meaningful.
As $\kappa$ approaches 3/2, the solution $y$ in panel (a) becomes very large.
The dotted line shows the approximate analytical solution of zeroth order \eqref{e29}. 
Although this analytical estimate remains above unity for all $\kappa$, the condition $y_0 \gg 1$ is fulfilled only for $\kappa \lesssim 3$. 
Only in this low-$\kappa$ range is the zeroth-order analytical approximation in good agreement with the exact numerical solution.
The solid line represents the refined analytical solution, the first-order approximation in \eqref{e39}, derived here for the first time and significantly improving the accuracy of the approximate solution for large $\kappa$.

Panel~(b) of Figure~\ref{fig1} presents the ES potential at the exobase $\Phi_E(r_0)$, as a function of $\kappa$. 
The exact solutions (orange squares) are obtained from~\eqref{e23} using the exact numerical evaluation of $y$ in panel (a). 
The zeroth order approximation \eqref{e31} (dotted line) reproduces the exact solution only for $\kappa \lesssim 3$, while at higher values it increasingly overestimates $\Phi_E(r_0)$.
The first-order analytical approximation (solid line) yields precise results for large values of $\kappa$.   
Panel~(c) shows the terminal SW velocity $V_{\mathrm{SW}}$, obtained from Eq.~\eqref{e5} using the values of $\Phi_E(r_0)$ from panel~(b). 
The first order approximation (solid line), again, works well over a wide range of $\kappa$, in contrast to the zero order approximation, which systematically overestimates $V_{\mathrm{SW}}$ at larger $\kappa$. 
From this panel, it follows that, by assuming that electrons at the exobase follow a SKD distribution, the observed SW speeds ($\approx300 - 800$ km/s) can be explained if their $\kappa \in (2.5,7)$; the smaller the $\kappa$, the faster the wind. 
This property of SKD models has previously been outlined (e.g., \citeA{Maksimovic_etal_1997})
, and our results are consistent with these findings. 
We emphasize that the first-order analytical solution, derived in Section~\ref{sec3_2} for the first time, provides an accurate estimate for both the ES potential at the exobase and the corresponding terminal SW speed, as shown in Fig.~\ref{fig1}.

\subsection{Radial profiles}
\label{sec3_5}

We now construct the radial profiles of the macro-parameters within the framework outlined in Section~\ref{sec:2_2}, using the boundary conditions at the exobase first provided by our exact (numerical) solutions. 
The same procedure can be performed with our approximate analytical expressions.
The density profile has the general form given in~\eqref{e7} and depends on the parameter $C_0$. 
Assuming Maxwellian protons at the exobase, as in \citeA{MeyerVernet-Issautier-1998}, the general definition of $C_0$, equation~\eqref{e8}, reduces to
\begin{align}\label{e40}
C_0 &= \frac{n_p(r_0)}{\sqrt{\pi}} \, P\!\left(\frac{V_{p0}}{w_{p0}}\right),
\end{align}
where
\begin{align} \label{e41}
{P(x)\over \sqrt{\pi}} =  \left( 1/2 - x^2 \right) e^{x^2} \big( 1 - \text{erf}(x) \big) + \frac{x}{\sqrt{\pi}}. 
\end{align}
$C_0$ incorporates the influence of the ES and gravitational potentials through $V_{p0}$, defined in \eqref{e9}.
To evaluate $V_{p0}$, we use the exact solution for $\Phi_E(r_0)$ (orange markers in Figure~\ref{fig1}~b) and the gravitational potential $\Phi_G(r_0)$ defined in~\eqref{e6}. 
The exobase altitude is specified as $r_0 = 6R_S$.

\begin{figure}[t!] 
\centering
\includegraphics[width=1\linewidth]{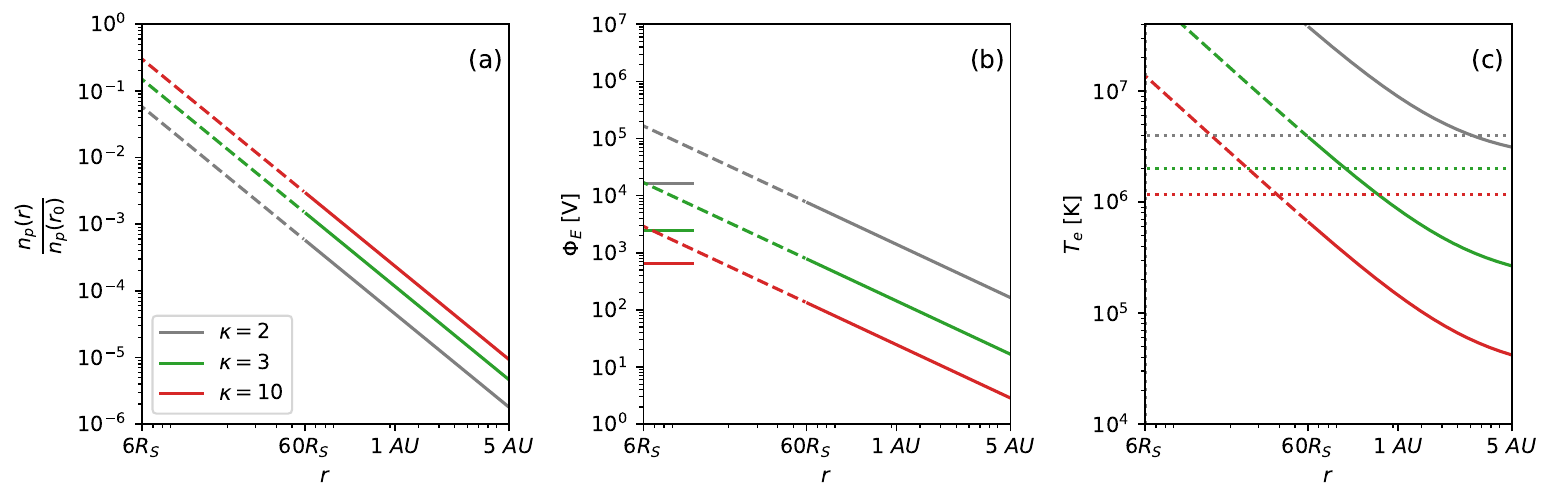}
\caption{Radial profiles for (a) normalized density, (b) ES potential, and (c) electron temperature as a function of distance \(r\) (in units of $R_{S}$). The curves of different colors correspond to different values of the parameter $\kappa = 2, 3, 10$. Short lines in panel (b) indicate the ES potential at the exobase $\Phi_E(r_0)$ (boundary value at $r_0=6R_S$). The dotted lines in panel (c) mark the assumed electron temperatures at the exobase (see text). 
}\label{fig2}
\end{figure}

The radial profile of the ES potential follows from the general equation~\eqref{e10},  
which depends on the parameter $D_0$. 
Using the general definition of $D_0$ in \eqref{e11}, where specified for SKD electrons, $D_0$ takes the form
\begin{align}\label{e42}
    D_0 &=\frac{2 n_{e0} A_{\kappa}}{3\big(\kappa \, w_{e0}^2 \big)^{3/2}} 
\left( 1 + y \right)^{-\kappa-1} 
\end{align}
where
\begin{align}\label{e43}
A_\kappa &= \frac{\Gamma(\kappa+1)}{\Gamma(\kappa - 1/2) \, \Gamma(3/2)},
\end{align}
and
\begin{align}\label{e44}
n_{e0} = 
\frac{n_e(r_0)}{1 - \tfrac{1}{2} \, A_\kappa \,
B_{x_0}\!\left(\kappa - \tfrac{1}{2}, \tfrac{3}{2}\right)}.
\end{align}
In \citeA{MeyerVernet-Issautier-1998} $n_{e0}$ is the normalization constant for the Kappa distribution, but here it is only a factor in expression \eqref{e42}.

The radial profile of the electron temperature is obtained from the general equations \eqref{e12}-\eqref{e14}, which involve $C_0$, $D_0$, and $B_2$. $C_0$ and $D_0$ are given by \eqref{e38} and \eqref{e42}, respectively, while $B_2$ follows from the general definition \eqref{e15}.  
For SKD electrons, it reduces to
\begin{align} \label{e45}
B_2 &= \frac{n_{e0} \, \kappa \, \theta_e^2}{2 (2\kappa - 3)} \, 
\left( 1 + y \right)^{-\kappa + \tfrac{1}{2}} \, 
\left( \kappa y + \tfrac{3}{2} \right),
\end{align}
where the exact solution $y$, obtained in Section~\ref{sec3_1}, is substituted.

Radial profiles are shown in Figure~\ref{fig2}. 
The gray, green and red curves correspond to $\kappa=2, 3, 10$, respectively. 
Since the model equations are derived in the limit of large distances ($\eta \equiv (r_0/r)^2 << 1$), we show the curves as solid lines only for $r/r_0>10$.
The segments at low radial distances (down to the exobase altitude), where this assumption is not satisfied, are shown as dashed lines to indicate their limited reliability.
Panel~(a) displays the density, normalized by its value at $r=r_0$. 
All profiles follow a power law dependence with slope $-2$. For smaller $\kappa$, the density is lower, although all curves, by definition, converge to unity at $r=r_0$.
This indicates that smaller $\kappa$ values correspond to steeper density gradients at short distances, where the model itself is not valid. 
From the continuity equation in spherically symmetric flow, we find $V_r(r)= \text{const} / (r^2n(r))$, such that a smaller $\kappa$ (lower $n$) implies faster acceleration, in accordance with previous results, e.g., \citeA{Maksimovic_etal_1997}. 
Panel~(b) of Figure~\ref{fig2} presents the radial profiles of $\Phi_E (r)$. The potential follows a universal power law and its magnitude is highly dependent on $\kappa$, when this parameter is small. Smaller $\kappa$ values, which indicate a higher proportion of suprathermal electrons, result in a larger (more positive) $\Phi_E$.
Panel~(c) shows the radial profiles of the electron temperature. 
At moderate but still large distances, temperature profiles decrease according to a power law, and then at larger distances they decrease asymptotically to a constant value. 
The temperatures at the exobase (dotted horizontal lines), $T_{\kappa 0} = \kappa/(\kappa -3/2) T_{e0}$, increase with decreasing $\kappa$, compared to the Maxwellian limit (large $\kappa$) $T_{e0}=10^6$~K.
For small $\kappa=2$ (gray lines), the temperature takes unrealistically high values that drop below the initial temperature only for vast distances.

\section{\bf RKD electrons and Maxwellian protons at the exobase}\label{sec4}
We now assume that the electrons at the exobase have an RKD distribution
\begin{align}\label{e46}
    f_{e0}=C_{RKD}\left(1+\frac{v^2}{\kappa w_{e0}^2} \right)^{-\kappa-1}\exp\left(-\frac{\alpha^2v^2}{w_{e0}^2}\right).
\end{align}
\corr{ The initial purpose of the cutoff parameter $\alpha$ was to ensure a consistent definition of all moments of the RKD model for any value of $\kappa > 0$ \cite{Scherer-etal-2017}. 
$\alpha$ can be determined from observations \cite{Scherer-etal-2021}, while the RKD model must reconcile two requirements: to reduce the non-physical effects of superluminal particles (with $v>c$, where $c$ is the speed of light in a vacuum) in SKDs, and at the same time to preserve the effects of the measured suprathermal tail \cite{Scherer-etal-2019}. 
As a result, this parameter must roughly satisfy $w_{e0}/c < \alpha < 1$, a condition that we will also take into account below when choosing values for $\alpha$ (for our $w_{e0} = 5.5 \times 10^3$~km/s, then $\alpha$ must satisfy $0.018 < \alpha <1$).}

The normalization constant $C_{RKD}$ is determined using \eqref{e3}, where the second integral on the right side is neglected in the limit of large $V_ >> w_{e0}$ \cite{MeyerVernet-Issautier-1998}
\begin{align}\label{e47}
    C_{RKD}=\frac{n_e (r_0)}{(\pi w_{e0}^2\kappa)^{3/2}U(3/2,3/2-\kappa,\alpha^2\kappa)},
\end{align}
where $U$ is the hypergeometric Kummer (or Tricomi) function \cite{Abramowitz-1965}.
For the electron flux in~\eqref{e1}, we obtain 
\begin{align}\label{e48}
F_e=&{\kappa^2 w_{e0}^4 \over 2} {C_{RKD} \;  \exp(-\beta y)} \times \left[\beta^{\kappa-1} U(\kappa,\kappa,\beta (1+y))-\beta^{\kappa}U(1+\kappa,1+\kappa,\beta (1+y))\right], 
\end{align}
where $\beta=\alpha^2\kappa$, and $y = V_0^2/(\kappa w_{e0})$ is the same as above in \eqref{e22}.
For Maxwellian protons, their flux on the right side of \eqref{e1} reads then explicitly as
\begin{align}\label{e49}
F_p=\frac{n_p(r_0) w_{p0}}{\pi^{3/2}}.
\end{align}
%

\subsection{Exact solutions} \label{sec4_1}
%
Substituting \eqref{e48} and \eqref{e49} into \eqref{e1}, and taking into account $n_e(r_0)=n_p(r_0)$, we find 
\begin{align}\label{e50}
\frac{w_{p0}}{w_{e0}}=\frac{\sqrt{\kappa}e^{-\beta y}}{2U(3/2,3/2-\kappa,\beta)}&[\beta^{\kappa-1}U(\kappa,\kappa,\beta(1+y))  -\beta^{\kappa}U(1+\kappa,1+\kappa,\beta(1+y))]. 
\end{align}
This equation is solved with Brent's root finding method \cite{Brent1973}, for the same thermal speeds at the exobase ($r_0 = 6~R_S$) $w_{p0} = 182$ km/s, and $w_{e0}=5.5 \times 10^3$ km/s, corresponding to $T_{p0}=m_pw_{p0}^2/2k_B=2 \times10^6$ K and $T_{e0}=m_ew_{e0}^2/2k_B= 10^6$ K.
Note that for RKD electrons, the physical temperature is a function of $\alpha$ and $\kappa$ \cite{Scherer_2019} 
\begin{align}\label{e51}
T_{r\kappa} =  T_{e0} \; \kappa \; U(5/2, 5/2-\kappa, \alpha^2 \kappa)/ U(3/2, 3/2-\kappa, \alpha^2 \kappa).
\end{align}
We have also examined the possibility of deriving an approximate analytical solution of \eqref{e50}. 
However, with the numerical solution for $y$ we found that the argument of the $U$-function, $z = \beta(1+y)$, remains close to unity. 
As a result, the $U$-function could not be expanded in a useful manner, preventing the derivation of an approximate analytical solution.

\begin{figure}[t!]
\centering
\includegraphics[width=0.99\linewidth]{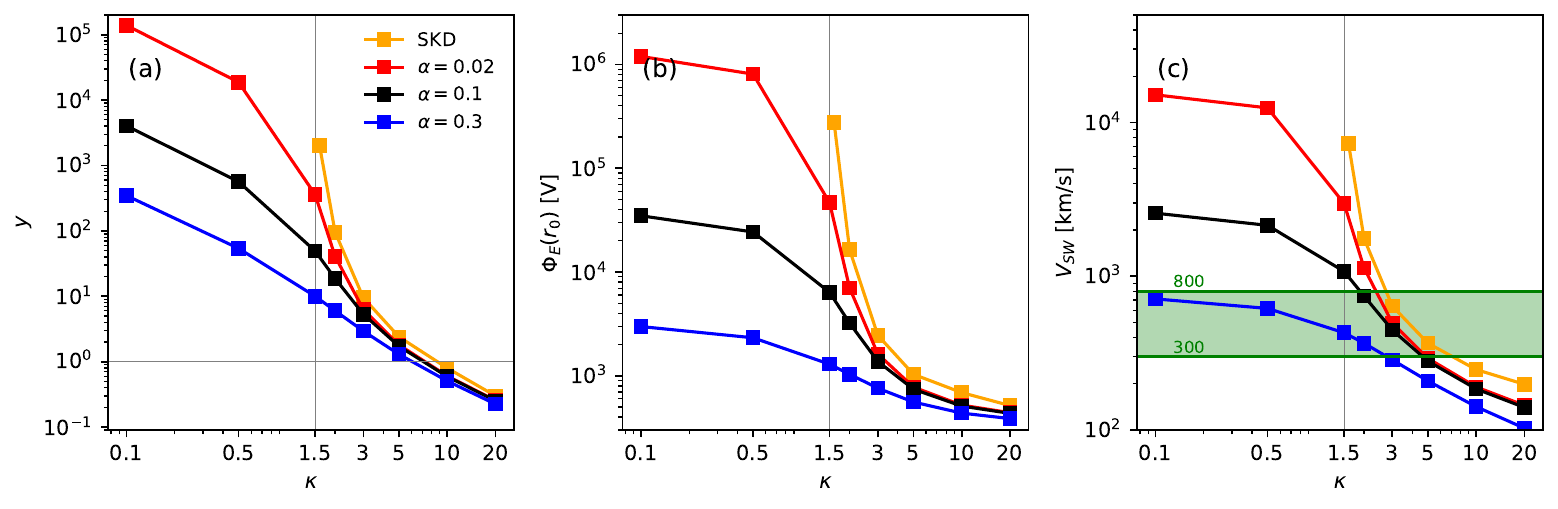}
\caption{The dependence of the (exact) solutions at the exobase on $\kappa$ for RKD electrons, in the same format as Fig.~\ref{fig1}. Results are plotted for three cutoffs $\alpha = 0.02$ (red), $\alpha = 0.1$ (black) and $\alpha = 0.3$ (blue), and are compared with the results for SKD electrons (orange) from Fig.~\ref{fig1}. The greenish range in panel (c) indicates observed SW speeds (300–800 km/s). 
} \label{fig3}
\end{figure}

The exact solutions for $y$ are presented in Figure~\ref{fig3}, panel (a), with the blue, black, and red curves corresponding to the RKD electrons with different values of the cutoff parameter $\alpha$. 
For comparison, the orange curve shows the exact solution of \eqref{e24} for SKD electrons. 
As the cutoff parameter $\alpha$ decreases to 0, the RKD reduces to the SKD, and the RKD solutions become closer to those for SKD electrons.
Panel~(b) of Figure~\ref{fig3} shows the ES potential at the exobase $\Phi_E(r_0)$, derived from $y$ in panel~(a), using \eqref{e23}.
Panel~(c) of Figure~\ref{fig3} represents the SW terminal velocity $V_{SW}$ obtained from \eqref{e5}, using $\Phi_E(r_0)$ from panel~(b).
At large $\kappa$, the $y(\kappa)$ curves for RKD electrons with different $\alpha$ converge very closely. 
However, the corresponding curves obtained for $\Phi_E(r_0)$ and $V_{SW}$ are still slightly separated; see the slight difference between the red (RKD solution for lowest (still finite) $\alpha = 0.02$) and orange curves. 
This is mainly because the relationship between $y$ and $\Phi_E$ involves a scaling factor, linearly dependent on $\kappa$, which amplifies even minor differences when $\kappa$ increases.
In addition, the derivation of \eqref{e13} for SKD accounts for both integrals in \eqref{e3}, without assuming that $V_0$ is much larger than the most probable speed of the electron distribution. 
In contrast, the RKD derivation of equation \eqref{e50} relies on the approximation that $V_0$ is much larger than the most probable electron speed, allowing the second integral in \eqref{e3} to be neglected. 
This simplification likely accounts for the remaining mismatch.

The most important feature of the model with RKD electrons at the exobase is the possibility of investigating extended conditions, including the low-$\kappa \leqslant 3/2$ regimes. 
The applicability of SKD is limited to $\kappa>3/2$.
We find that for RKD electrons, $\Phi_E(r_0)$ and the resulting $V_{SW}$ approach finite values as $\kappa\to 0$, with the saturation level controlled by $\alpha$. For SKD electrons, reproducing the observed SW speed (green band) requires $3<\kappa<7$. 
In contrast, RKD electron models can predict, depending on the values of the parameter $\alpha$, a bulk velocity $V_{SW}$ in the observationally relevant range even for very small $\kappa$, below the critical value for SKD, $\kappa \leqslant 3/2$.

\subsection{Radial profiles}
\label{sec4_2}

We now investigate how regularization of the suprathermal distribution tails (quantified by the parameter $\alpha$) modifies the radial profiles compared to the SKD model.
We keep the assumption that protons follow a Maxwellian distribution, as in Section~\ref{sec3}. 
In the general expression \eqref{e7} for the radial density profile, we now use $C_0$ given by \eqref{e40} and \eqref{e41}. 
The difference arises in the evaluation of $V_{p0}$ from \eqref{e9}: instead of using the ES potential at the exobase derived for SKD electrons, we now adopt $\Phi_E(r_0)$, shown in Figure~\ref{fig3}~(b) and obtained for RKD electrons.
The radial profile of the ES potential is determined by the general equation \eqref{e10}, which involves two parameters $C_0$, as explained above, and $D_0$. 
The parameter representing the contribution of the non-escaping electrons, $D_0$, is evaluated using \eqref{e11}, 
where $f_{e0}$ is the RKD electron distribution function defined in \eqref{e46}. 
To build the radial profile of the electron temperature, we use the general equations \eqref{e12}-\eqref{e14}. 
Parameters $C_0$ and $D_0$ are calculated as explained above, while $B_2$ is given by
\begin{align}\label{e54}
    B_2=\pi \int_{V_0}^{\infty} dv v^3 \sqrt{v^2-V_0^2}f_{e0}(v),
\end{align}
and is evaluated numerically.
Physically, the product $B_2 \eta$ represents the contribution of escaping electrons to the second-order velocity moment, i.e., their role in determining the temperature profile.

\begin{figure}[t]
\centering
\includegraphics[width=1.0\linewidth]{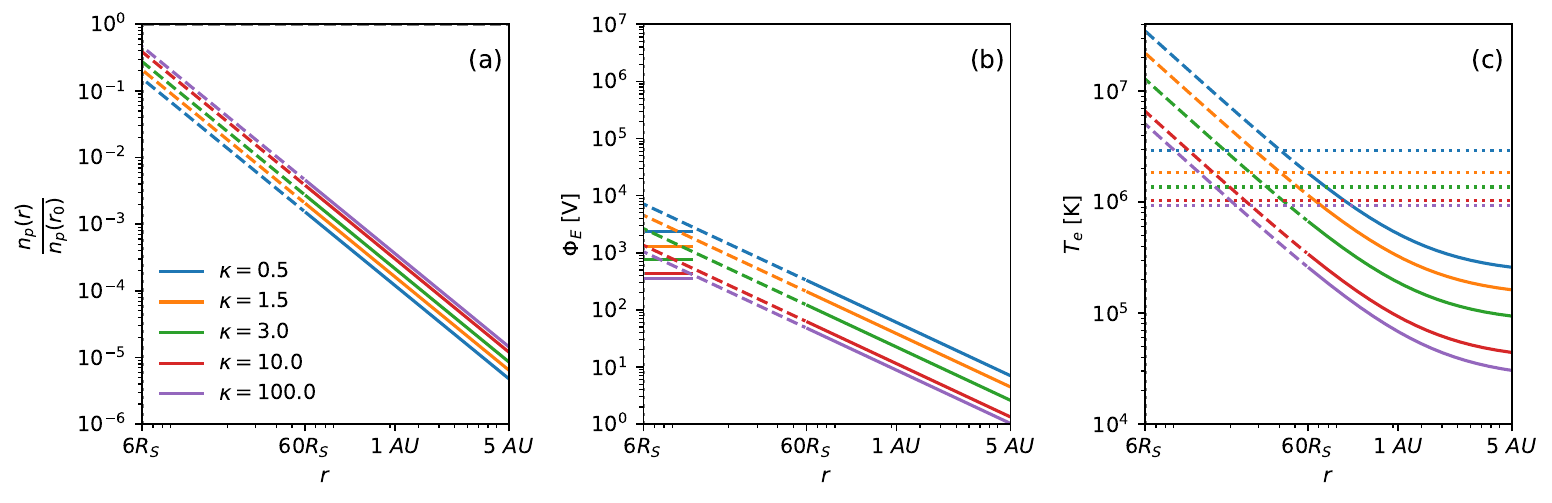}
\caption{Radial profiles for (a) normalized density, (b) ES potential and (c) electron temperature for RKD electrons with $\alpha=0.3$ and different values of $\kappa$. Short lines in panel (b) indicate $\Phi_E(r_0)$ (boundary values) at the exobase $r_0 = 6R_S$). The dotted lines in panel (c) mark the assumed electron temperatures at the exobase (see text). }
\label{fig4}
\end{figure}

Figure~\ref{fig4} shows the resulting radial profiles for RKD electrons with a strong cut-off parameter $\alpha=0.3$ and different values of $\kappa$. 
Radial profiles are obtained in the limit of large distances from the Sun ($\eta = r_0^2/r_2 << 1$).
As in Figure~\ref{fig2}, the profiles are shown as solid lines for $r>60 R_S$, where this assumption is valid, and with dashed lines for lower distances.
Panel~(a) of Figure~\ref{fig4} shows the density profiles normalized to their value at $r_0$. 
The density is lower for smaller $\kappa$. This trend is the same as in the case of SKD electrons, shown in Figure~\ref{fig2}(a). 
However, in contrast to the case with SKD electrons, the densities for different $\kappa$ are much closer. 
Again, by the definition of normalization, all profiles should converge to 1 at $r=r_0$. 
Thus, the density gradient at short distances (where the model is not valid) is smaller in the model with RKD electrons, and the SW acceleration is reduced. 
The reason is that $\alpha=0.3$ imposes a strong cut-off that significantly suppresses the suprathermal distribution tails.
Panel~(b) of Figure~\ref{fig4} shows the radial profiles of the ES potential. 
$\Phi_E$ increases, as expected if suprathermal electron tails become heavier (smaller $\kappa$). 
In contrast to the case with SKD, shown in Figure~\ref{fig2}~(b), $\Phi_E(r)$ increases only slightly as $\kappa$ decreases. 
The reason for this effect is that the cut-off parameter reduces the unrealistic part of the tails of the SKD electron distribution.
Panel~(c) of Figure~\ref{fig4} shows the radial profiles of the electron temperature, with dotted lines marking the initial RKD electron temperatures given by \eqref{e51}; see also Table~\ref{tab2} for the conversion factors obtained for different values of $\alpha$ and $\kappa$. 
The overall trend resembles that of the SKD case in Figure~\ref{fig2}~(c): smaller values of $\kappa$ correspond to higher $T_e$. However, unlike SKD, the RDK with $\alpha = 0.3$ predicts lower temperatures, even for small $\kappa$, which may provide more realistic descriptions.

\begin{figure}[t]
\centering
\includegraphics[width=1.\linewidth]{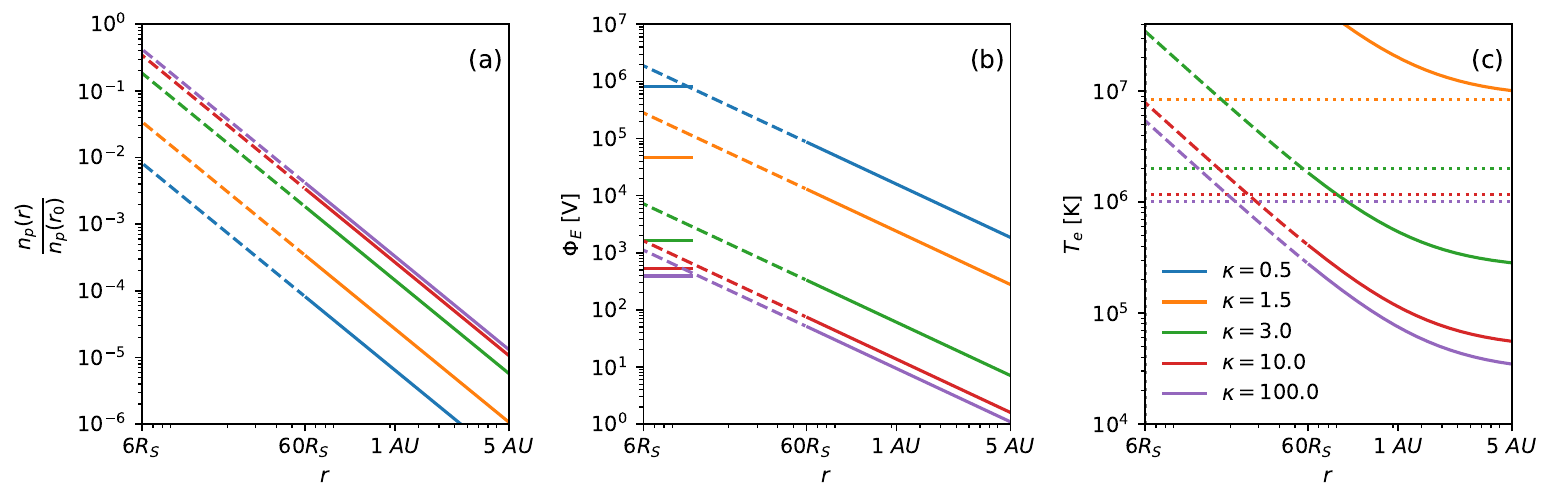}
\caption{Radial profiles of (a) normalized density, (b) ES potential and (c) electron temperature for the model with RKD electrons with $\alpha=0.02$ and different values of $\kappa$. Short lines in panel (b) indicate $\Phi_E(r_0)$ (boundary values) at the exobase $r_0 = 6R_S$). The dotted lines in panel (c) mark the assumed electron temperatures at the exobase (see text).}
        \label{fig5}
\end{figure}

Figure~\ref{fig5} shows the radial profiles for RKD electrons with a much smaller cut-off parameter $\alpha = 0.02$. 
In this limit, the profiles are more similar to those obtained for SKD electrons, as shown in Figure~\ref{fig2}.
However, the RKD formulation also allows exploration of $\kappa \leqslant 3/2$, corresponding to even heavier-tailed distributions and not accessible in the SKD model. 
In particular, for these new cases, the relative density shown in panel~(a) drops to very low levels at large distances, implying a stronger SW acceleration.
As an immediate consequence, in panel~(b), the ES potential reaches higher values. 
Also, the temperature shown in panel~(c) increases to higher values (e.g., for $\kappa < 3$), which are unrealistic for the SW, but can eventually describe stellar winds of stars with much hotter coronas \cite{Guedel-2004}. 
The contrasting results obtained in Figures~\ref{fig4} and \ref{fig5} demonstrate the important role of the regularization parameter $\alpha < 1$.
Despite the complexity of the new RKD models, they can describe a more pronounced presence of suprathermal electrons even in the solar corona and a consistent contribution to the acceleration of SW streams, without overestimating the acceleration due to unphysical (superluminal) populations. 
\corr{It is also very plausible that excessive suprathermal electrons with $\kappa \leqslant 3/2$ can be found in plasma sources of energetic events, such as coronal eruptions and coronal mass ejections (CMEs). 
If future in situ observations, or at least remote spectral analyses \cite{Dudik-2021, Effenberger-2021}, adapt RKDs and highlight energetic electrons at the exobase, then exospheric models could explain CME-like outflows with speeds of up to several thousand km/s.}

\begin{table}[h]
\centering
\caption{Conversion factors $T_{rk}/ T_{e0}$ relating Maxwellian temperature ($T_{e0} = 10^6$ K) to RKD temperature $T_{rk}$ defined by \eqref{e51}, for various values of $\alpha$ and $\kappa$.} \label{tab2}
\begin{tabular}{c c c}
\toprule
$\alpha$ & $\kappa$ & \text{$T_{rk}/ T_{e0}$} \\
\midrule
0.02 & 0.5   & 226.89 \\
0.02 & 1.5   & 8.41 \\
0.02 & 3.0   & 1.99 \\
0.02 & 10.0  & 1.18 \\
0.02 & 100.0 & 1.01 \\
\hline
0.3  & 0.5   & 2.90 \\
0.3  & 1.5   & 1.84 \\
0.3  & 3.0   & 1.37 \\
0.3  & 10.0  & 1.04 \\
0.3  & 100.0 & 0.93 \\
\bottomrule
\end{tabular}
\end{table}
%

\section{Conclusions}

Recent advances in both observations of suprathermal electrons in the solar corona and consistent modeling of these suprathermal populations using RKDs give new impetus to macroscopic modeling of the SW based on the kinetic properties of its components, electrons and protons.
RKDs are adjustments to SKDs to address the latter's inconsistencies (see details in the introduction).
In this paper, we have exploited the method proposed by \citeA{MeyerVernet-Issautier-1998} for the semi-analytical derivation of key parameters at the exobase, such as the ES potential, as well as the properties of the SW, e.g., density, electron temperature, and their variation with heliocentric distance (radial profiles).
First, we have refined the previous model based on SKD electrons at the exobase, for which we derived rigorous, fully analytical approximations.
These facilitate not only numerical calculations in the development of exospheric models but also comparison with the new estimates based on the presence of RKD electrons at the exobase.

New models with RKD electrons can prevent overestimating SW properties at large distances, including energies, temperatures, and bulk speeds. 
\corr {Recent PSP observations \cite{Abraham-etal-2022, Zheng-etal-2024} reveal an increased abundance of suprathermal electrons at the exobase associated with reduced $\kappa$ parameters, e.g., $\kappa \to 4$ (and perhaps even less, according to the decreasing trend), for which SKD modeling generally leads to such overestimations.}
The advantage of RKD models is that the suprathermal influence can be weighted by adjusting the new cutoff parameter $\alpha < 1$.
Thus, estimates of the temperatures and bulk speeds of the SW can remain realistic, depending on the value of $\alpha$, and be of the order of those reported by observations.
In this case, the new parameterizations are still at a semi-analytical level, due to their complexity, being expressed in terms of Tricomi hypergeometric integral functions that do not allow for physically meaningful approximations.
Instead, an extensive and consistent characterization is offered for various initial (or boundary) conditions at the exobase.
Electron VDs with strong suprathermal tails described by low $\kappa \leqslant 3/2$, i.e., below the critical threshold for SKDs, can also be considered. 
These can be associated, as a plausible hypothesis (still to be confirmed by the observations), with events much more energetic than the SW, such as coronal flares and CMEs.
In addition, such conditions cannot be excluded as being conducive to the modeling of stellar winds in the astrospheres of stars with much hotter coronas, with temperatures up to two orders of magnitude higher than that of the solar corona \cite{Guedel-2004}.

\appendix
\section{Moments for protons} \label{secA}
In the limit of large distances, the proton moments of order $q$ can be approximated as \cite{MeyerVernet-Issautier-1998}
\begin{align}\label{Ae1}
    M_{pq}\approx C_q \eta,
\end{align}
where
\begin{align}\label{Ae2}
    C_q=\pi\int_{0}^{\infty}dv\frac{v^3}{(v^2+V_{p0}^2)^{(1-q)/2}}f_{p0}(v),
\end{align}
$V_{p0}$ is defined by \eqref{e9}, and $f_{p0}$ is the proton distribution function at the exobase.
The derivation uses conservation of energy and magnetic moment, along with Liouville's theorem. 
The phase-space limits are simplified to calculate the moments, assuming large distances from the Sun.

\section{Moments for electrons}\label{secB}
In the limit of large distances, the moment of order $q$ of the electron distribution can be approximated as follows \cite{MeyerVernet-Issautier-1998}
\begin{align}\label{Be1}
    M_{eq}\approx D_q v_{M}^{q+3}+B_q \eta,
\end{align}
where 
\begin{align}\label{Be2}
D_q=4\pi f_{e0}(V_0)/(q+3),
\end{align}
and
\begin{align}\label{Be3}
B_q=\pi\int_{V_0}^{\infty}dv v^3 (v^2-V_0^2)^{(q-1)/2} f_{e0} (v).
\end{align}
The first and second terms in \eqref{Be1} correspond to nonescaping and escaping electrons, respectively.
%

\section{Equivalence of normalizations}\label{secC}

Here we demonstrate that the SKD normalization used in \citeA{MeyerVernet-Issautier-1998} is equivalent to that used in section~\ref{sec3}.
Meyer-Vernet defines the distribution with a generic constant $n_{e0}$, which is related to the physical density at the exobase $n_e(r_0)$ through their 
Eq.~(63). 
This relation involves the function $\beta_2(y)$ defined in their
Eq.~(64). 
In contrast, our formulation, given in \eqref{e19} uses the incomplete beta function $B_{x_0}(3/2, \kappa - 1/2)$ with $x_0 = y/(1+y)$.
The link between these two forms follows from the \emph{symmetry relation} for the incomplete beta function 
\cite[equation 6.6.3]{Abramowitz-1965}
\begin{equation}
B_x(a,b) + B_{1-x}(b,a) = B(a,b).
\end{equation}
Applying this identity with $z = 1/(1+y)$ yields
\begin{equation}
B_{x_0}\!\left(\tfrac{3}{2}, \kappa - \tfrac{1}{2}\right) 
= \frac{1 - \beta_2(z)}{A_\kappa},
\end{equation}
where $A_\kappa$ is given in \eqref{e43}.

\section*{\corr{Conflict of Interest Statement}}
\corr{The authors have no conflicts of interest to disclose.}

\section*{Open Research Section}
%
\corr{
The Jupyter Notebooks to execute the analysis in the paper can be found at
\url{https://github.com/ISashaVinogradov/kappa-electrons-solar-wind},
is hosted at GitHub, and is preserved at Zenodo
\cite{vinogradov_2026_github_kappa} (Version JGR2025\_v1.0,
\url{https://doi.org/10.5281/zenodo.18152868}). }
%
%
Solutions were obtained using Brent’s method (brentq) from the publicly available SciPy library.  
We also used mpmath (BSD license) to evaluate special functions with arbitrary precision, and NumPy for numerical array operations.  
The plots were generated using Matplotlib, which is distributed under the Matplotlib (PSF) license available at \url{https://matplotlib.org/}.

\acknowledgments
The authors acknowledge support from the Ruhr-University Bochum and the Katholieke Universiteit Leuven. 
These results were also obtained in the framework of the projects C16/24/010 (C1 project Internal Funds KU Leuven), G0B5823N (FWO-Vlaanderen), WEAVE project G002523N / FI~706/31-1 (FWO-Vlaanderen / DFG-Germany), and 4000134474 (SIDC Data Exploitation, ESA Prodex). Powered@NLHPC: This research was partially supported by the supercomputing infrastructure of the NLHPC (CCSS210001). The resources and services used in this work were provided by the VSC (Flemish Supercomputer Center), funded by the Research Foundation - Flanders (FWO) and the Flemish Government.
SP is funded by the European Union (ERC, Open SESAME, 101141362). Views and opinions expressed are, however, those of the author(s) only and do not necessarily reflect those of the European Union or the European Research Council. Neither the European Union nor the granting authority can be held responsible for them.

\bibliography{References}

\end{document}